\documentclass[runningheads]{cl2emult}

\usepackage{makeidx}  
\usepackage{graphicx} 
\usepackage{subeqnar} 
\usepackage{multicol} 
\usepackage{cropmark} 
\usepackage{eso}      
\makeindex            

\newcommand{\euler}[1]{{\usefont{U}{eur}{m}{n}#1}}

\newcommand{\umu}{\mbox{\euler{\char22}}}

\begin{document}
\title*{ISO Observations of Star-forming Galaxies}
\toctitle{ISO Observations of Star-forming Galaxies}
%
%
\titlerunning{ISO Observations of Star-forming Galaxies}
%
\author{Bahram Mobasher\inst{1}
\and Jos\'e Afonso\inst{2}
\and Lawrence Cram\inst{3}}
\authorrunning{B. Mobasher et al.}
%
%
\institute{Space Telescope Science Institute, 3700 San Martin Drive, 
\\Baltimore MD 21218, USA -- also affilliated with the Space Sciences Department of the European Space Agency
\and Blackett Laboratory, Imperial College of Science, Technology and Medicine,
Prince Consort Road, London SW7 2BW, England, UK
\and School of Physics, University of Sydney, Sydney, NSW 2006, Australia}

\maketitle              

\begin{abstract}

The Infrared Space Observatory {\it (ISO)} is used to carry out 
mid-IR (7 and 15 \umu m) and far-IR (90 \umu m) observations
of a sample of star-forming sub-mJy radio sources. By selecting
the sample at radio wavelengths, one avoids biases due to
dust obscuration. It is found that the mid-IR luminosities, 
covering the PAH features, measure the star formation rate for galaxies 
with $P_{1.4 GHz} < 10^{23}$ W Hz$^{-1}$. This is further confirmed using 
the H$\alpha$ luminosities. The far-IR emission is also found to 
trace the SFR over the whole range of radio and H$\alpha$ luminosities. 
The implication of the mid-IR measurements in estimating the SFRs 
from the future infrared space missions (SIRTF and ASTRO-F) is discussed.

\end{abstract}

\section{Introduction}
There now exist several measurements of the star formation rate (SFR) 
at different redshifts, based on UV~\cite{1,2,3} and 
Balmer-line~\cite{4,5,6,7}
studies, with the latter yielding estimates a factor of 2--3 times higher 
than the former, presumably because of differential dust extinction.  
These disagreements impede progress in understanding the evolution 
with redshift of the rates of star-formation and heavy element 
production~\cite{8}. 
The problem becomes more serious at high redshifts due to changes in dust 
content in galaxies with look-back time.  In particular, optically selected 
samples are likely to be biased against actively star-forming and dusty 
galaxies, leading to an underestimation of the SFR from these samples.  
Indeed, it has been shown that a large fraction of the bolometric luminosity 
emerges at far--IR wavelengths, with recent observations with the
Infrared Space Observatory (ISO) showing that
the contribution to the cosmic infrared background is dominated by
infrared luminous galaxies. This confirms that most of the star formation, 
specially at high redshifts, is hidden in dusty environments.
Also, it is shown that different star-formation diagnostics give
different SFRs even for the same galaxy. Therefore, to
accurately trace the SFR, one needs to use as many {\it independent} 
star-formation diagnostics as possible.

In this study, the sensitivity of the mid-IR fluxes (7-15 \umu m), 
covering the PAH features, to the star-formation activity in galaxies
will be studied, using
an unbiased sample of star-forming galaxies. The potential of this
technique in measuring the SFRs at $z\sim 2$ is then discussed.

\section{Sample Selection}

The sample for this study consists of sub-mJy radio sources, selected at 
radio (1.4\,GHz) wavelengths~\cite{9} and hence, is   
free from dust-induced selection biases. 
A total of 400 of these galaxies are then spectroscopically 
observed with their redshifts measured and spectral features 
(H$\alpha$, MgII, etc) identified~\cite{10}. 
A sample of 65 radio sources were then observed with ISOCAM 
(7 and 15 \umu m) and ISOPHOT (90 \umu m)-
(Afonso et al 2001, {\it in preparation}).  
The objects adopted for {\it ISO} observations are chosen to be sub-mJy
radio sources, showing evidence for star-formation activity in their spectra
and sufficiently bright at mid- to far- IR wavelengths (as predicted from
their SEDs) to allow detection at these wavelengths. 
The number of radio sources in the {\it ISO} survey region, together with
the number of galaxies with detections 
at the three {\it ISO} wavelengths are listed in Table~\ref{tab1}. 
The ISOCAM pointed survey
also resulted in the serendipitous detection of 26 sources for which no radio 
counterpart was found. These objects will not be discussed here. 
Details about the {\it ISO} observations and data reduction will be presented
in a future paper (Afonso et al 2001, {\it in preparation}). 

\begin{table}
\centering
\caption{Number of sources in the the areas covered by both the
{\it ISO} and radio surveys. 
$N_s$ and $N_d$ denote, respectively, 
the number of radio sources over the area covered by the {\it ISO}
(65 pointings for ISOCAM,
and 44 for ISOPHOT) and the number of {\it ISO} detected sources.}
\renewcommand{\arraystretch}{1.4}
\setlength\tabcolsep{5pt}
\begin{tabular}{rrr}
\hline\noalign{\smallskip}
Band & $N_s$ & $N_d$ \\
\noalign{\smallskip}
\hline
\noalign{\smallskip}
 7 \umu m & 146  & 16 \\
15 \umu m & 146 &  15 \\
90 \umu m &  44 &   9 \\
\hline
\end{tabular}
\label{tab1}
\end{table}

\section{Results}

The intrinsic luminosities at the {\it ISO} and radio wavelengths are estimated
assuming $H_0 =65$ km/sec/Mpc. The K-corrections are applied, assuming
a flat spectrum at 7 and 15 \umu m wavelengths. For the 90 $\mu$m and 1.4\,GHz 
fluxes, a  power-law SED of the form $f_\nu \propto \nu^{n}$ is assumed
with spectral indices of respectively $n=-2$ and $-0.7$. 

\begin{figure}[t]
\centering
\includegraphics[width=.9\textwidth]{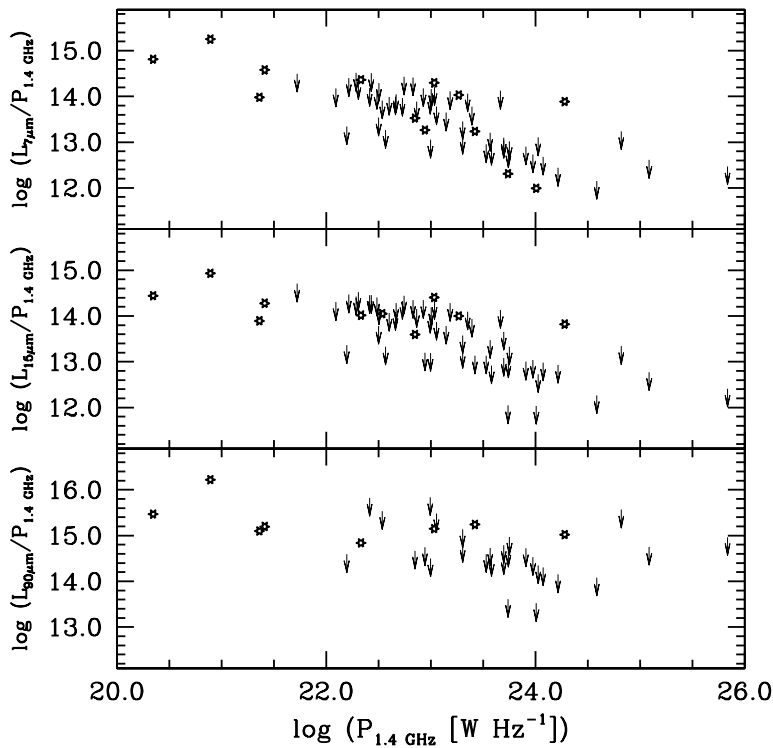}
\caption[]{Ratio of the {\it ISO} luminosities to the radio power 
as a function of the radio (1.4\,GHz) power. $L_{7\umu m}$, $L_{15\umu m}$ and $L_{90\umu m}$ are defined as $\nu P_\nu$ at the respective rest-frame wavelength and are given in units of $L_\odot$.}
\label{fig1}
\end{figure}

The ratio of the {\it ISO} (7, 15, 90 \umu m) to 
radio power as a function of the radio power for galaxies
in the present sample is shown in Fig.~\ref{fig1}. Both the detections 
and upper limits are included in this diagram.  Figure~\ref{fig1} is
significant in that,  the lack of a trend here
indicates that the radio, compared to mid-IR and far-IR luminosities measure 
the {\it same}
quantity (ie. star-formation) whereas, the presence of a trend implies that
they are sensitive to {\it different} physical processes.

The $L_{7\umu m}/P_{1.4\,{\rm GHz}} - P_{1.4\,{\rm GHz}} $ and 
$L_{15\umu m}/P_{1.4\,{\rm GHz}} - P_{1.4\,{\rm GHz}}$ relations both show 
a slight trend for
log$(P_{1.4\,{\rm GHz}}) < 10^{23}$\,W/Hz, followed by a steep slope at
log$(P_{1.4\,{\rm GHz}}) > 10^{23}$\,W/Hz. The value of $10^{23}$\,W/Hz 
corresponds
to the characteristic radio power of the sub-mJy sources 
where also a change of slope is found in the 1.4\,GHz luminosity
function of star-forming galaxies~\cite{13}. 
Assuming that 
the radio emission from galaxies is a measure of the synchrotron radiation 
due to
relativistic electrons, produced by supernovae remnants, 
and hence their SFR~\cite{11,12}, one concludes that for
$ P_{1.4\,{\rm GHz}} <  10^{23}$\,W/Hz, 
the mid-IR (7 and 15 \umu m) 
luminosity is sensitive to the star-formation activity.  
However, for objects with $ P_{1.4\,{\rm GHz}} > 10^{23}$ W/Hz, 
the PAH molecules are destroyed due to the strength of the photon field, 
resulting a decrease in the mid-IR flux from galaxies. 
At the far-IR 90 \umu m wavelength, there is no significant trend on the
 $L_{90\umu m}/P_{1.4\,{\rm GHz}} - P_{1.4\,{\rm GHz}}$ diagram, 
confirming that both
the far-IR and radio luminosities measure the same quantity (ie. SFR).
These results are obtained using both the detections and upper limits.
Using only the detections, the trend in the relation disappears at 15 \umu m
while remains the same for 7 \umu m.

\begin{figure}[t]
\centering
\includegraphics[width=.9\textwidth]{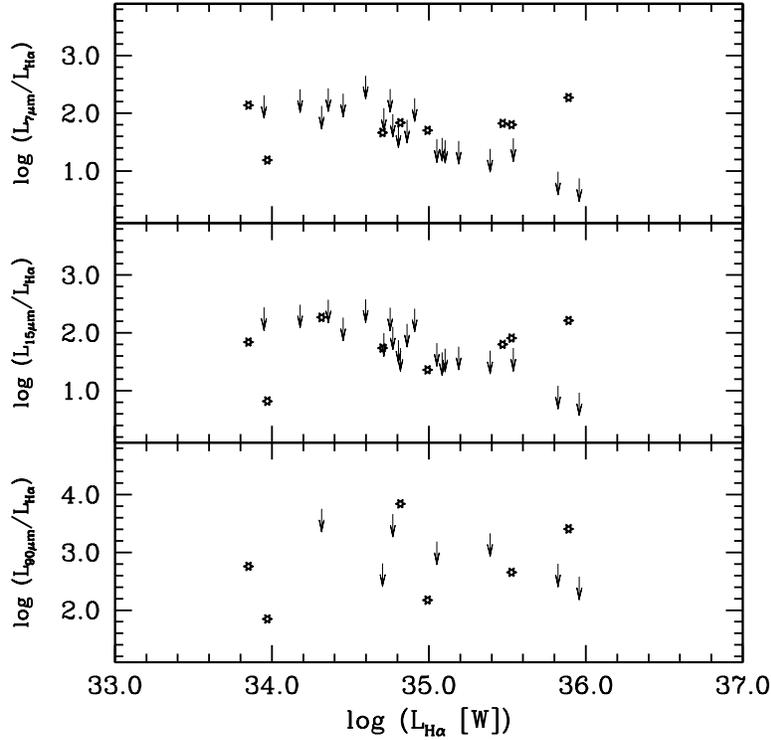}
\caption[]{Ratio of the {\it ISO} mid and far-IR luminosities to H$\alpha$
luminosity as a function of H$\alpha$ luminosity.}
\label{fig2}
\end{figure}

The above results are confirmed using H$\alpha$ line luminosity 
(Figure~\ref{fig2}) which is a more direct
measure of the star-formation in galaxies. 
While there is a small
trend on the $L_{7\umu m}/L_{H\alpha} - L_{H\alpha}$ relation
for $L_{H\alpha} > 10^{34.8}$ W, 
the trend almost disappears
for $L_{15\umu m}/L_{H\alpha} - L_{H\alpha}$ and is entirely absent on the   
$L_{90\umu m}/L_{H\alpha} - L_{H\alpha}$ relation. This
implies an increase in the sensitivity to the star-formation from 7 to 15 
and 90 \umu m wavelengths, in agreement with the results from Figure~\ref{fig1}. 

The results in this study can be used to establish new star-formation 
diagnostics, based on 7 and 15 \umu m luminosities, and to calibrate them.
This has significant
implications for future mid-IR surveys with the SIRTF and ASTRO-F. 
For example, a 24 \umu m deep
survey, planned with the SIRTF, can detect the rest-frame 7 and 15 \umu m 
emissions at $z= 2.4$ and $z=0.6$ respectively,  
allowing measurement of the SFRs between these redshifts. By selecting a sample
at radio wavelengths, with follow-up mid-IR observations, one could then 
avoid dust-induced selection biases in estimating the SFR.

\clearpage
\addcontentsline{toc}{section}{Index}
\flushbottom
\printindex


\begin{thebibliography}{7}
%
\addcontentsline{toc}{section}{References}

\bibitem{1} Lilly, S.~J., Le~Fevre, O., Hammer, F., \& Crampton, D. (1996)
The Canada-France Redshift Survey: the luminosity density and star 
formation history of the Universe to $z \sim 1$. ApJ, 460, L1
 
\bibitem{2} Treyer, M.~A., Ellis, R.~S., Milliard, B., Donas, J., \& 
Bridges, T.~J. (1998) An Ultraviolet-Selected Galaxy Redshift Survey: 
new estimates of the local star formation rate. MNRAS, 300, 303
 
\bibitem{3} Connolly, A.~J., Szalay, A.~S., Dickinson, M., Subbarao, M.~U., \&
Brunner, R.~J. (1997) The Evolution of the Global Star Formation History 
as Measured from the Hubble Deep Field. ApJ, 486, L11
 
\bibitem{4} Gallego, J., Zamorano, J., Aragon-Salamanca, A., \& Rego, M. (1995)
The Current Star Formation Rate of the Local Universe. ApJ, 455, L1
 
\bibitem{5} Pettini M. et al, 1997, ASP conference series
 
\bibitem{6} Tresse, L., \& Maddox, S.~J. (1998) The H$\alpha$ Luminosity 
Function and Star Formation rate at $z \sim 2$. ApJ, 495, 691
 
\bibitem{7} Glazebrook, K., Blake, C., Economou, F., Lilly, S., \& Colless, M. 
(1999) Measurement of the Star Formation Rate from H$\alpha$ in field Galaxies 
at $z=1$. MNRAS, 306, 843
 
\bibitem{8} Ellis R., (1998) The formation and evolution of galaxies. 
Nature, 395, 3 
 
\bibitem{9} Hopkins, A.~M., Mobasher, B.,
 Cram, L., \& Rowan-Robinson, M. (1998) The PHOENIX Deep Survey: 
1.4-GHz source counts. MNRAS, 296, 839

\bibitem{10} Georgakakis, A., Mobasher, B.,
 Cram, L., Hopkins, A., Lidman, C., \& Rowan-Robinson, M. (1999) 
The PHOENIX Survey: optical and near-infrared observations of 
faint radio sources. MNRAS, 306, 708

\bibitem{11} Cram, L., Hopkins, A.,
 Mobasher, B., \& Rowan-Robinson, (1988) Star Formation Rates in Faint 
Radio Galaxies. ApJ, 507, 155

\bibitem{12} Condon, J.~J. (1992)
Radio emission from normal galaxies. ARA\&A, 30, 575

\bibitem{13} Mobasher, B., Cram, L., Georgakakis, \& A., Hopkins, A. (1999)
The 1.4\,GHz and H$\alpha$ Luminosity Functions and Star Formation
Rates from Faint Radio Galaxies. MNRAS, 308, 45

\end{thebibliography}
\end{document}